\documentclass[journal,article,submit,pdftex,moreauthors]{Definitions/mdpi} 
\firstpage{1} 
\pubvolume{1}
\issuenum{1}
\articlenumber{0}
\pubyear{2022}
\copyrightyear{2022}
\datereceived{} 
\dateaccepted{} 
\datepublished{} 
\hreflink{https://doi.org/} % If needed use \linebreak
\usepackage{algorithmic} % 算法相关格式
\usepackage{algorithm}
\usepackage{float}

\makeatletter	% This code is used to solve the subfigure compatibility problem
\let\c@lofdepth\relax
\let\c@lotdepth\relax
\makeatother
\usepackage{subfigure}
\Title{A Primal-Dual Based Power Control Approach for Capacitated Edge Servers}

\TitleCitation{A Primal-Dual Based Power Control Approach for Capacitated Edge Servers}

\Author{Qinghui Zhang $^{1}$\orcidA{}, Weidong Li $^{2}$, Qian Su $^{1}$ and Xuejie Zhang$^{1}$*}

\AuthorNames{Qinghui Zhang, Weidong Li, Qian Su and Xuejie Zhang}

\AuthorCitation{QH, Z.; WD, L.; Q, S.; XJ, Z.}
\address{%
$^{1}$ \quad School of Information Science and Engineering, Yunnan University, Kunming, Yunnan 650091, P.R. China; zhangqinghui@mail.ynu.edu.cn, suqian@ynu.edu.cn, xjzhang@ynu.edu.cn\\
$^{2}$ \quad School of Mathematics and Statistics, Yunnan University, Kunming, Yunnan 650091, P.R. China; weidong@ynu.edu.cn}

\corres{Correspondence: xjzhang@ynu.edu.cn}

\abstract{The intensity of radio waves decays rapidly with increasing propagation distance, and an edge server's antenna needs more power to form a larger signal coverage area. Therefore, the power of the edge server should be controlled to reduce energy consumption. In addition, edge servers with capacitated resources provide services for only a limited number of users to ensure the quality of service (QoS). We set the signal transmission power for the antenna of each edge server and formed a signal disk, ensuring that all users were covered by the edge server signal and minimizing the total power of the system. This scenario is a typical geometric set covering problem, and even simple cases without capacity limits are NP-hard problems. In this paper, we propose a primal-dual-based algorithm and obtain an $m$-approximation result. We compare our algorithm with two other algorithms through simulation experiments. The results show that our algorithm obtains a result close to the optimal value in polynomial time.}


\begin{document}

%%%%%%%%%%%%%%%%%%%%%%%%%%%%%%%%%%%%%%%%%%
\section{Introduction}
\subsection{Background}
% ???????,????????????,????????????????????????????????????????????????????????????????????????????????????,??????????????????????????????????2010??2020?????,global internet traffic has expanded 15-fold,???????????????????????????????????????????????????,???????????????????????????????????
%Despite exponential growth in demand for these services, rapid improvements in energy efficiency have helped to limit electricity demand growth.

In recent years, edge computing has been proposed as a timely and resource-efficient alternative to address data computation issues \cite{Abbas2018}. Edge computing brings the service and utilities of cloud computing closer to the user. The response time that users perceive is effectively reduced, and the data processing in the cloud center is alleviated. In addition to reduced service delays, we must consider another issue. From 2010 to 2020, global Internet traffic expanded 15-fold, and the energy consumed in transmitting data has increased at the same rate \cite{IEA2021}. Industry and academia have focused on reducing energy consumption in wireless communication processes, in which the power of the antenna is an important consideration.
% ??????WSN????????
%Various power control methods have been proposed for different wireless networks. 
Various different power control methods have been proposed for wireless networks. In terms of controlling the power of cellular networks, \cite{DBLP:journals/corr/abs-2205-00429} interpreted cellular and cell-free massive MIMO networks as max-min utility optimization problems with affine interference mappings and polyhedral constraints. In \cite{Dai2021}, Dai \emph{et al.} investigated the joint optimization of base station (BS) clustering and power control for non-orthogonal multiple access (NOMA)-enabled coordinated multipoint (CoMP) transmission in dense cellular networks, maximizing the sum rate of the system.
In addition, in terms of wireless sensor networks (WSNs), \cite{Chincoli2018} investigated how machine learning could be used to reduce the possible transmission power level of wireless nodes and, in turn, satisfy the quality requirements of the overall network. Reducing the transmission power has benefits in terms of both energy consumption and interference. In \cite{DBLP:journals/tcom/MoltafetLCP22}, Moltafet \emph{et al.} developed a dynamic control algorithm using the Lyapunov drift-plus-penalty method. They used this approach to jointly optimize the sampling action of each sensor, the transmit power allocation, and the subchannel assignment, minimizing the average total transmit power of all sensors, subject to a maximum average Age of Information (AoI) constraint for each sensor.

% ?????????????????
% ????????????WSN??????????,????????,?????????????????????,??????,????????????,??CPU??????
%The above research on cellular networks and WSNs controlled the antenna's power by reducing signal interference and improving the user's quality of service (QoS).
The above researches optimized the antenna power by considering the influence of each user on the system and ensuring the user’s quality of service (QoS).
QoS may be data transmission rate, response time or task execution time, etc.
However, in edge networks, some computing resources, such as CPUs, are often scarcer than channel resources. Edge networks often have no remaining CPU resources before they run out of channel resources.
Therefore, wireless channel allocation and signal interference are not considered in this paper.
% 此外，有向天线信号在当今的蜂窝通信中使用的十分广泛。但基于MIMO技术的蜂窝网络，可以把一个基站看做一个全向的信号发射设施。在本文中，我们把基站看做一个能发射全向信号的天线。
In addition, directional signal of antenna is very commonly used in today's cellular communication. In this paper, we abstract the MIMO-based base station as an omni-directional antenna. So the signal coverage of the antenna can be seen as a disk.

% ??????????????????????,??????????????????????????????????????,??????server????????????????,????????????????????????????????,??????????????????????????????????????,???????????????????????????,??????????????????????????,???????????????????,???????????????????,?????????????????IP????????????IP??????????????????????,the problem studied in this article is how to get a power assignment for capacitated edge servers so that them can cover all users and minimize the total power.
% ???,??????,?????????????????:1.?MEC???????????????(??????????MPC??)?2.???NP-Hard?,???PD???3.(Theoretical analysis),????,???????????4.performance evaluation, ????,??????????
In such an edge networks based on the above assumptions, the signal coverage area of an antenna, which is determined by the antenna's power, is a disk centered on the it. Because the signal intensity decreases with increasing distance, a larger signal coverage disk needs more power \cite{Ronnow2019}. In an edge computing system composed of edge servers and users, the edge servers communicate with users through wireless signals transmitted by antennas. This article assumes that the edge server has omni-directional antenna and is abbreviated as \emph{server}. When a server serves a user, its signal should cover the user. In addition, each edge server can provide only limited resources for the users it serves. In summary, the problem studied in this article is how to appropriately assign power in a capacitated edge server to ensure that the server covers all users while minimizing the total power.
We define this problem as the \emph{capacitated minimum power cover} (CMPC) problem. 
%In practical situations, a user may be located outside the maximum coverage region of the servers or the number of users may exceed the total capacity of the servers. We assume that a server has no upper power limit and that the overall system has the capacity to cover all users. As a result, we can obtain a scheme that covers all users with the technique developed in this paper, which provides a reference for later decisions, such as whether to cover distant users. We can also address the situation in which the demand exceeds the capacity by repeatedly using the method proposed in this paper.Therefore, the above two assumptions are reasonable.  
The signal-to-noise ratio (SNR) between the server and the user is usually the a factor affecting the quality of communication. We know that SNR is inversely proportional to the signal distance. We make $SNR_{min}$ to be the SNR between a server $i$ and $j$ user, where the distance between them is the farthest in system. We assume that as long as the SNR between any server and user is greater than $SNR_{min}$, the server could serve this user. In addition, the overall system has the capacity to cover all users. As a result, we can obtain a approximate scheme that covers all users with the technique developed in this paper, and explain the extreme case where a server serves the furthest user rarely occurs. We can also address the situation in which the demand exceeds the capacity by repeatedly using the method proposed in this paper. Therefore, the above two assumptions are reasonable.

Our detailed contributions in this work can be summarized as follows:
\begin{itemize}
	\item \emph{(CMPC and Resource Allocation Model)}
	The CMPC problem is a fundamental \emph{minimum power coverage} (MPC) problem. The MPC problem is NP-hard even in the absence of capacity constraints \cite{Alt2006, Bilo2005}. Based on the CMPC problem, we establish a minimum power control resource allocation model in a capacitated edge network.
	\item \emph{(Primal-Dual Algorithms)}
	To address the above challenges, we propose a primal-dual-based approximation algorithm to solve the CMPC problem. After the theory proof, we obtain an approximation guarantee of $m$ ($m$ edge servers considered) in polynomial time.
	\item \emph{(Performance Evaluation)}
	Numerical results are presented to validate the effectiveness and efficiency of our proposed algorithms.
\end{itemize}

\subsection{Previous Research}
%MEC???Primal-dual?????????
The primal-dual-based approximation algorithm is a generalization of the primal-dual method used for linear programming and combinatorial optimization problems \cite{DavidP.2011}. This algorithm provides effective solutions for many optimization problems in mobile edge computing (MEC). In \cite{9199545}, attack-resilient distributed algorithms based on primal-dual optimization were proposed for situations when Byzantine attackers are present in a system. In \cite{9653102}, Wang \emph{et al.} modelled the system in an online manner and formulated the underlying optimization problem, maximizing the total profit according to constraints on the computational resources on the edge clouds and job migration. Furthermore, a (1 - 1/e)-competitive primal-dual-based online algorithm was proposed. An efficient social welfare approximation algorithm that used a classic primal-dual framework was developed in \cite{DBLP:journals/ton/ZhouLWH17}. In this paper, the cloud market for computing jobs with completion deadlines was studied, and efficient online auctions for cloud resource provisioning were designed.

%??????
Broadly, the CMPC problem belongs to the family of \emph{minimum weight set cover} (MWSC) problems. 
In general, MWSC problem is challenging to solve optimally, even for some simple versions. For example, Alt \emph{et al.} and Bil\`{o} \emph{et al.} presented a minimum cost covering problem without capacity constraints that is still NP-hard for any $\alpha > 1$ in \cite{Alt2006, Bilo2005}. Thus, the development of polynomial-time approximation algorithms is the main objective for CMPC problems. In \cite{Qinghui2022}, Zhang \emph{et al.} proposed a local-ratio based power control approach for access point in mobile edge computing. About the theoretical study of the minimum power cover (MPC) problem, Liu \emph{et al.} introduced  the $k$-prize-collecting minimum power cover problem ($k$-PCPC) where $k$ is the number of users need be covered in \cite{Liu2021}. They presented a novel two-phase primal-dual algorithm for the $k$-PCPC with an approximation ratio of at most $3^{\alpha}$. In recent work \cite{Liu2022}, Liu \emph{et al.} considered MPC problem with submodular and linear penalties. For the minimum power partial cover problem, Dai \emph{et al.} present an $O(\alpha)$-approximation algorithm in \cite{Dai2022_1}.
% 引用能量覆盖的相关研究，包括应用：本人local-ratio、无人机信号覆盖；理论：代涵、李伟东等人成果；rouding算法那篇

About disk cover problem, some researchers focused on minimizing the cardinality of disks. In \cite{Lyu2017}, Lyu \emph{et al.} considered UAV-Mounted mobile base stations (UAV-MBS)  with the same radii to provide wireless coverage for a group of distributed ground terminals and minimize the number of UAV-MBS. In \cite{Bandyapadhyay2020}, a rounding-based mechanism for capacitated covering problems that minimized the cardinality was proposed and obtained a constant approximation to address this problem.
In the past, there have been some researches on the minimum weight disk cover problem.
Varadarajan considered the weight \cite{Varadarajan2010} and presented a clever quasi-uniform sampling technique that was improved by Chan \emph{et al.} \cite{Chan2012}, yielding a constant approximation for the minimum weight disk cover problem. This constant approximation was generalized by Bansal and Pruhs \cite{Bansal2012} for the minimum weight disk multicover problem in which every point must be covered multiple times.

%In previous studies, the capacity constraints of the minimum power cover problem have rarely been considered. In \cite{Bandyapadhyay2020}, a rounding-based mechanism for capacitated covering problems that minimized the cardinality and obtained a constant approximation was proposed to address this problem. However, the radius of the disk was expanded to reduce the cardinality. Thus, this method is not applicable in this paper due to the exponential relationship between radius and power.

Energy-efficient optimization has attracted extensive attention in mobile computing. By endowing edge servers multiple power states, e.g., active state and sleep state, it is promising to improve the total energy consumption of edge servers through switching under-utilized servers into sleep state in \cite{Wang2019}. Ali \emph{et al.} in \cite{Zaiwar2019} proposed a novel energy-efficient deep learning based offloading scheme (EEDOS) to select an optimal set of computation components to offload to ESs, aiming at minimizing the energy consumption of MDs.  Li et al. in \cite{Li2018AnET} studied the channel selection for task offloading. The effect of multi-channel interference on the energy efficiency of task offloading was taken into account.
was taken into account. Gu \emph{et al.} in reference \cite{Gu2018} studied the problem of how to efficiently assign computing tasks to reduce energy consumption in the edge computing system under the constraints of the computing capacity of both MDs and ENs, wireless channel conditions, and delay.
%一些关于边缘计算中能耗优化的文章。联合考虑延迟为优化目标的等等。

\section{System Model and Preliminaries}

\subsection{System Model}
We consider that all the facilities in the edge networks are distributed in a fixed dimensional ${\mathbb{R}^d}$ space, where the facilities are edge servers and users. Let $S$ denote the set of $m$ edge servers and $U$ denote the set of $n$ users. For each user $j \in U$ to connect to an edge server $i \in S$, $j$ must be contained in the signal disk formed by $i$ and obtain an IP from this disk. The IP capacity of edge server ${i}$ is $k_i$. If $i$ forms a signal disk with radii $r_i$, the power we should provide is
\begin{equation}
	\label{eq:power}
	{p_i} = c \cdot {({r_i})^\alpha },
\end{equation}
where $c$ and $\alpha$ are constants ($\alpha$ is usually called the \emph{attenuation factor}).

Although $r_i$ can be selected arbitrarily, there will be precisely one terminal device located on the boundary of the disk with radius $d(i, j)$ in the optimal solution, where $d(i, j)$ denotes the distance between $i$ and $j$. Therefore, at most $mn$ disks need to be considered. We denote the set of such disks as $\mathcal{D}$. $i$ and $j$ can form a disk $D_{ij} \in \mathcal{D}$ with radius $r_{ij} = d(i, j)$ and center $i$. The disk set with $i$ as the center in $\mathcal{D}$ is represented by $\mathcal{D}_i$. Therefore, we have $\mathcal{D}_i = n$. A set of disks $\widehat {\mathcal{D}} \in {\mathcal{D}}$ is called a cover for a set of users $\widehat U \in U$ if each user $j \in \widehat U$ is covered by some disk in $\widehat {\mathcal{D}}$. The problem is to find a cover $\widehat {\mathcal{D}} \in {\mathcal{D}}$ for $S$ with minimum power $p(\widehat {\mathcal{D}}) = \sum\nolimits_{{D_{ij}} \in \widehat {\mathcal{D}}} {{p_{ij}}} $.

In the following, we denote instance of the CMPC problem as $\left({U,{\mathcal{D}},k} \right)$ and the optimal power for the instance $\left({U,{\mathcal{D}},k} \right)$ as $OPT$. To simplify the notation, we use $D_{ij}$ to represent both a disk in $\mathcal{D}$ and the set of users contained in $D_{ij}$ and $p_{ij}$ to denote the power of disk $D_{ij}$, where ${p_{ij}} = c \cdot {({r_{ij}})^\alpha }$.
We note that $D_{ij}$ containing a user $u$ means that $u$ is within the range of $D_{ij}$ ($u\in D_{ij}$); thus, $D_{ij}$ covering $u$ means that server $i$ serves $u$, who then obtains resources. 
In addition, we use $SNR_{min}$ denote the SNR between a server-user pair $(s,u)$ where $(s,u) = \arg \max \;d(i,j)$. Therefore, the SNR between any other server-user pair will not be less than $SNR_ {min}$ and a server could serve any users. Obviously, the power of $s$ occurred by serving $i \in U\backslash \{ u\} $ would not more than the power by serving $u$.

In actual scenarios, there are situations in which multiple users are located at the same distance from a server. Suppose that a server cannot provide services for all users simultaneously. In that case, the server provides the users with services in an orderly manner according to the urgency or bid level of the users. Because this article does not focus on task scheduling, we use Definition 1 to determine the order of the users.

\begin{Definition} 
	\label{d1}
	Map the positions of all users and servers to a coordinate system and use $\cos({\overrightarrow {su} })$ to represent the cosine of the angle between vector $\overrightarrow {su}$ and the $x$-axis formed by server $s$ and user $u$. The distances between users $u$ and $d$ and server $s$ are $r_{su}$ and $r_{sd}$, respectively. If $\cos (\overrightarrow {su}) > \cos (\overrightarrow {sd})$, then $r_{su} \succ r_{sd}$, and the same is true if $p_{su} \succ p_{sd}$. Because $r_{su} = r_{sd}$, ${D_{su}}$ contains $d$ but ${D_{sd}}$ does not contain $u$.
\end{Definition}

\begin{table}[]
	\caption{Summary of Notations}
	\centering
	\label{table:notation}
	\begin{tabular}{|ll|ll|}
		\hline
		\multicolumn{1}{|l|}{\cellcolor[HTML]{FFFFFF}{\color[HTML]{333333} $m$}} & \# of servers & \multicolumn{1}{l|}{$n$} & \# of users \\ \hline
		\multicolumn{1}{|l|}{$S$}   & Set of servers & \multicolumn{1}{l|}{$U$} & Set of users            \\ \hline
		\multicolumn{1}{|l|}{$U'$}  & Set of uncovered users & \multicolumn{1}{l|}{$\mathcal{D}'$} & Set of unselected disks\\ \hline
		\multicolumn{1}{|l|}{$k_i$}     & \multicolumn{3}{l|}{Server $i$'s capacity}                            \\ \hline
		\multicolumn{1}{|l|}{$d(i, j)$}  & \multicolumn{3}{l|}{Distance between server $i$ and user $j$}       \\ \hline
		\multicolumn{1}{|l|}{$D_{ij}$}  & \multicolumn{3}{l|}{Disk formed by $i$ and $j$ or the set of users the disk contains}\\ \hline
		\multicolumn{1}{|l|}{$r_i$ ($r_{ij}$)}     & \multicolumn{3}{l|}{Radius of server $i$'s disk (disk $D_{ij}$)}         \\ \hline
		\multicolumn{1}{|l|}{$p_i$ ($p_{ij}$)}     & \multicolumn{3}{l|}{Power of server $i$'s disk (disk $D_{ij}$)}         \\ \hline
		\multicolumn{1}{|l|}{$\mathcal{D}$}     & \multicolumn{3}{l|}{ Set of disks formed by $m$ servers and $n$ users}   \\ \hline
		\multicolumn{1}{|l|}{$\mathcal{D}_i$}     & \multicolumn{3}{l|}{ Set of disks centered on server $i$ in $\mathcal{D}$}  \\ \hline
		\multicolumn{1}{|l|}{$x_{ij}$}     & \multicolumn{3}{l|}{Disk $D_{ij}$ is selected (1) or not (0)}         \\ \hline
		\multicolumn{1}{|l|}{$y_{h, ij}$}     & \multicolumn{3}{l|}{User $h$ is covered by disk $D_{ij}$ (1) or not (0)}     \\ \hline
		\multicolumn{1}{|l|}{$\theta_h$}     & \multicolumn{3}{l|}{Costs charged due to user $h$}         \\ \hline
		\multicolumn{1}{|l|}{$\mu_i$}     & \multicolumn{3}{l|}{Extra cost of server $i$ selecting multiple disks}         \\ \hline
		\multicolumn{1}{|l|}{$\beta_{ij}$}& \multicolumn{3}{l|}{Minimum cost that disk $D_{ij}$ charges each user it contains}\\ \hline
		\multicolumn{1}{|l|}{$\gamma_{h, ij}$}& \multicolumn{3}{l|}{Cost that user $h$ is willing to pay for disk $D_{ij}$}\\ \hline
		\multicolumn{1}{|l|}{$D'_{ij}$}& \multicolumn{3}{l|}{Set of uncovered users in $D_{ij}$}\\ \hline
		\multicolumn{1}{|l|}{$\delta(D_{ij})$}& \multicolumn{3}{l|}{Set of users charged by $D_{ij}$}\\ \hline
	\end{tabular}
\end{table}

\section{A Primal-Dual Approach for Capacitated Servers}
In this section, we present a primal-dual algorithm for the CMPC problem on the instance $(U,\mathcal{D}, k)$. Then, we show how to use this algorithm to determine the power assignment of each server.

\subsection{Capacitated Minimum Power Cover Problem}
%% ???????,????????????,????????????????????
Under the assumption of sufficient capacity ($\sum\nolimits_{i \in S} {{k_i}} \ge n$), the CMPC problem can be formulated as an integer program. Variable $x_{ij}$ indicates whether disk $D_{ij} \in \mathcal{D}$ is chosen, that is, $x_{ij} = 1$ if and only if $D_{ij}$ is selected. Variable $y_{h, ij}$ indicates whether user $h \in U$ is covered by $D_{ij}$; here, $y_{h, ij} = 1$ if and only if $h$ is covered by $D_{ij}$. The integral linear programming (ILP) problem can be formulated as follows:

\begin{align}
	\min  & \sum\limits_{{D_{ij}} \in \mathcal{D}} {{p_{ij}} \cdot {x_{ij}}} \label{ip:2}\\
	s.t. &  \sum\limits_{{D_{ij}}:h \in {D_{ij}}} {{y_{h,ij}}}  \ge 1,h \in U\;{\text{and}}\;{D_{ij}} \in {\mathcal{D}} \tag{\ref{ip:2}{a}}\label{ip:2.1}\\
	&  {k_{i}}{x_{ij}} - \sum\limits_{h:h \in {D_{ij}}} {{y_{h,ij}}}  \ge 0,{D_{ij}} \in \mathcal{D} \tag{\ref{ip:2}{b}}\label{ip:2.2}\\
	&  {x_{ij}} \ge {y_{h,ij}}, h \in {D_{ij}} \in \mathcal{D} \tag{\ref{ip:2}{c}}\label{ip:2.3}\\
	&  \sum\limits_{{D_{ij}} \in {\mathcal{D}_i}} {{x_{ij}}}  \le 1,{\mathcal{D}_i} \in \mathcal{D} \tag{\ref{ip:2}{d}}\label{ip:2.4}\\
	&  {x_{ij}} \in \left\{ {0,1} \right\},{y_{h,ij}} \in \left\{ {0,1} \right\},\forall h \in U,\forall {D_{ij}} \in \mathcal{D} \tag{\ref{ip:2}{e}}\label{ip:2.5}%
\end{align}

Note that constraint (\ref{ip:2.1}) ensures that user $h$ is covered by at least one disk. The capacity limit of each disk is expressed in constraint (\ref{ip:2.2}). Constraint (\ref{ip:2.3}) guarantees that disk $D_{ij}$ can not cover user $h$ until the disk is selected. 
Constraint (\ref{ip:2.4}) implies that each server $i \in S$ can select only a disk as its power assignment.

The ILP without constraints (\ref{ip:2.2} and \ref{ip:2.4}) is still an NP-hard combinatorial optimization problem that is equivalent to the classic set cover problem. The challenge escalates when we consider the capacity of the servers and the uniqueness of the power assignments. To address these challenges, we utilize the primal-dual algorithm design technique. We relax the ILP constraints of $x_{ij}$ and $y_{h, ij}$ to $x_{ij} \ge 0$ and $y_{h, ij} \ge 0$ to formulate the dual problem. Note that we do not need to add the constraints $x_{ij} \le 1$ and $y_{h, ij} \le 1$ since they are automatically satisfied in an optimal solution of Equation (\ref{ip:2}). By introducing dual variables ${{\theta _h}}$, ${\beta _{ij}}$, ${\gamma _{h,ij}}$ and ${{\mu _i}}$ to constraints (\ref{ip:2.1}), (\ref{ip:2.2}), (\ref{ip:2.3}) and (\ref{ip:2.4}), respectively, the dual LP of the relaxed (\ref{ip:2}) becomes:
\begin{align}
	\max & \sum\limits_{h \in C} {{\theta _h}}  - \sum\limits_{i \in S} {{\mu _i}} \label{ip:3},\\
	s.t. & {\theta _h} - {\beta _{ij}} - {\gamma _{h,ij}} \le 0,\;h \in {D_{ij}} \in {\mathcal{D}} \tag{\ref{ip:3}{a}}\label{ip:3.1},\\
	& {k_{i}}{\beta _{ij}} + \sum\limits_{h:h \in {D_{ij}}} {{\gamma _{h,ij}}} \le {p_{ij}} + {\mu _i},{D_{ij}} \in {\mathcal{D}_i} \in \mathcal{D} \tag{\ref{ip:3}{b}}\label{ip:3.2},\\
	& {\theta _h} \ge 0,{\beta _{ij}} \ge 0,h \in U,{D_{ij}} \in {\mathcal{D}} \tag{\ref{ip:3}{c}}\label{ip:3.3},\\
	& {\gamma _{h,ij}} \ge 0,h \in {D_{ij}} \in {\mathcal{D}} \tag{\ref{ip:3}{d}}\label{ip:3.4}.%
\end{align}
These dual variables also have economic benefits. Servers charge a fee of $\theta_h$ to user $h \in U$ to provide services. The dual variable ${{\mu _i}}$ represents the additional cost when server $i \in S$ selects multiple disks. Users are not charged more than they are willing to pay. For a user $h$ contained in $D_{ij}$, $h$ is willing to pay $\beta_{ij}$ when $D$'s capacity is insufficient. Otherwise, the fee paid by $h$ is $\gamma_{h,ij}$. Furthermore, all users contained in disk $D_{ij}$ are willing to pay no more than the sum of $p_{ij}$ and $\mu_i$. Therefore, the objective function (\ref{ip:3}) maximizes the profit of the servers.

We next design an efficient primal-dual covering scheme that simultaneously increases the dual variables by a polynomial number of times, which we use to solve optimization problems (\ref{ip:2}) and (\ref{ip:3}).

\subsection{Primal-Dual Algorithm Design}
The first use of the primal-dual-based approximation algorithm is based on the work of Bar-Yehuda and Even \cite{Bar-Yehuda1981}. The procedure $PD$ follows the classic primal-dual method: starting from the trivial dual feasible solution of zero, the method increases the dual variables simultaneously until some disk becomes tight. Then, a tight disk is chosen and iterated until a feasible solution is obtained. Next, we introduce how the algorithm works in detail.

Initially, the dual variables $\{\theta\}$ and $\{\alpha\}$ are 0, resulting in a dual feasible solution (with all $\beta_{ij}=0$ and $\gamma_{h,ij}=0$). We use $U'$ to denote the set of uncovered users and $\mathcal{D}'$ to denote the set of unselected disks. We use $D'_{ij}$ to denote the uncovered users currently in $D_{ij}$. The disks are selected by increasing the dual variables $\theta_h$ for the uncovered users $h \in U'$ simultaneously. The dual program has two kinds of constraints: user constraints and server constraints.

To maintain the dual feasibility of the user constraints (\ref{ip:3.1}), as we increase $\theta_h$, we must increase $\beta_{ij}$ or $\gamma_{h,ij}$, where $h \in {D_{ij}} \in {\mathcal{D}}'$. If the disk contains a large number of uncovered users, we increase $\beta_{ij}$; otherwise, we increase $\gamma_{h,ij}$. Formally, if ${k_i} < \left| {D{'_{ij}}} \right|$, we increase $\beta_{ij}$; otherwise, we increase $\gamma_{h,ij}$.

For each disk constraint, ${k_i}{\beta _{ij}} + \sum\nolimits_{h:h \in {D_{ij}}} {{\gamma _{h,ij}}} \le {p_{ij}} + {\mu _i}$; initially, the left-hand side of this equation is 0, and the right-hand side is equal to the cost of the disk. This algorithm ensures that each server selects at most one disk; thus, $\mu_i = 0$ for all $i\in S$. When the dual variables of the unassigned edges are increased, we stop the procedure as soon as a disk constraint is met with equality. (In Algorithm \ref{alg:PD}, this is represented by disk $D_{su}$ in the main loop.) We can confirm only that the users in $D'_{su}$ are served by $s$, not that $s$ chooses disk ${D_{su}}$. We temporarily select this disk as the current disk for $s$ and record the value with $L_s=D_{su}$. Through $L_s$, we can ensure that each edge server selects at most one disk as its power strategy, so ${\mu _i} = 0, i \in S$ in the whole algorithm process. In lines 5-6 of Algorithm \ref{alg:PD}, we update the related variables and sets. First, we delete the users in $D'_{su}$ from $U'$ and the concentric disk with a radius less than $d(s, u)$ from $\mathcal{D}'$ (${\mathcal{D}}{'_{ < {r_{su}}}}=\{ {D_{ij}}:{D_{ij}} \in {\mathcal{D}}'\;\text{and}\; {r_{ij}} < {r_{su}}\} \cup {D_{su}}$). Then, we update $\{D'\}$ in the disks of the other servers. After these variables and sets are updated, the dual variables corresponding to the removed users and disks stop increasing in subsequent iterations. 
The above steps are iterated until all users are covered by a disk. The disk that covers the user are the last disk selected by each server, that is, $\{L\}$.

The above process maintains dual feasibility. According to Lemma \ref{lemma1}, the capacity constraint is maintained throughout the algorithm.

\begin{figure}[!t]
	
	\renewcommand{\algorithmicrequire}{\textbf{Input:}}
	\renewcommand{\algorithmicensure}{\textbf{Output:}}
	\begin{algorithm}[H]
		\caption{PD}
		\begin{algorithmic}[1]\label{alg:PD}
			\REQUIRE A set of users $U$, a set of servers $S$, a power function $p:{D_{ij}} \mapsto {\mathbb{R}^ + }$, and a capacity constraint $k$.
			\ENSURE A subset of disks $\widehat {\mathcal{D}}$ covering all users in $U$.
			\STATE Initialize $k{'_i} \leftarrow {k_i}, {L_i} \leftarrow \emptyset ,\;{\mu _i} \leftarrow 0,{\theta _h} \leftarrow 0,\;{\beta _{ij}} \leftarrow 0,\;{\gamma _{h, ij}} = 0,\;\forall j, h \in U,\;\forall i \in S$.
			\STATE $\widehat {\mathcal{D}} \leftarrow \emptyset ,\;{\mathcal{D}}' \leftarrow {\mathcal{D}},\;U' \leftarrow U$.
			\WHILE{$U' \ne \emptyset$}
			\STATE Increase ${\left\{ {{\theta _h}} \right\}_{h \in U'}}$ and ${{\text{\{ }}{\beta _{ij}} + {\gamma _{h,ij}}{\text{\} }}_{D{'_{ij}}:h \in D{'_{ij}}}}$ simultaneously until some disk $D_{su}$ becomes tight. (If $\left| {\left\{ {h:h \in D{'_{ij}} \cap U'} \right\}} \right| > {k_{i}}$, we increase $\beta_{ij}$; otherwise, we increase $\gamma_{h, ij}, h \in D{'_{ij}}$)
			\STATE ${L_s} \leftarrow {D_{su}}$.
			\STATE $U' \leftarrow U'\backslash D{'_{su}},{\mathcal{D}}' \leftarrow {\mathcal{D}}'\backslash {\mathcal{D}}{'_{ < {r_{su}}}}.$
			\STATE $D{'_{ij}} \leftarrow D{'_{ij}}\backslash \{ D{'_{ij}} \cap D{'_{su}}\} ,{D_{ij}} \in {\mathcal{D}}'\backslash {\mathcal{D}}{'_s}$.
			\ENDWHILE
			\STATE $\widehat{\mathcal{D}} \leftarrow \{ L\}$.
			\RETURN $\widehat{\mathcal{D}}$.
		\end{algorithmic}
	\end{algorithm}
\end{figure}

\subsection{A PD Instance }
\begin{figure}[htbp]
	\centering
	\includegraphics[width=6cm]{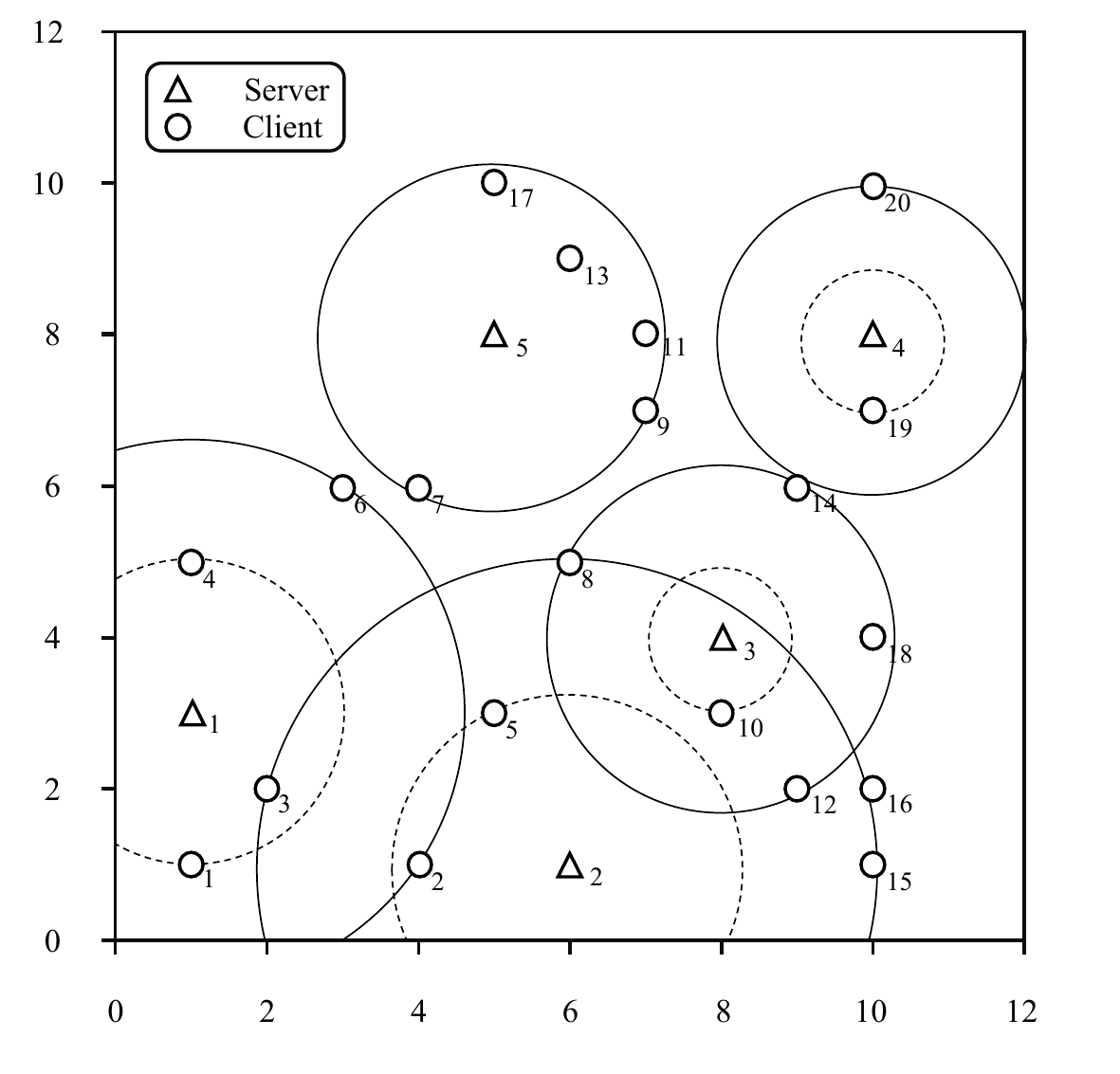}
	\caption{Total power of the PD algorithm with different numbers of servers and values of $K$.}
	\label{fig:instance}
\end{figure}

To further understand the PD algorithm, we illustrate an instance of the PD algorithm. In Fig. \ref{fig:instance}, we present an instance of the CMPC problem $\left({U,{\mathcal{D}},k} \right)$, where $U = \{ 1,2,...,20\} $, $S = \{ 1,2,3,4,5\} $ and $k=5$. For ease of representation, we use triangles and circles to represent servers and users, respectively, which are distributed in a 2-dimensional coordinate system. In Fig. \ref{fig:instance}, the disk drawn as a solid line is the final disk obtained by Algorithm \ref{alg:PD}. The specific power value that each server should provide can be calculated with the equation $p(a) = c \cdot r{(a)^\alpha }$ (in this case, $c=1$ and $\alpha=2$). The disk drawn with the dotted line is the disk that is temporarily selected by the PD algorithm during processing (that is, the value assigned to the variable $l_a$ during the algorithm's execution). There are nine disks in Fig. \ref{fig:instance}, indicating that the main loop of the PD algorithm produces nine tight disks in total. Next, we introduce how the PD algorithm obtained the results in this instance.

% ??????????,??D3,10????????10???????3??? According to the iteration sequence, the disks selected for each iteration are ${D_{3,10}}$, ${D_{4,19}}$, ${D_{5,7}}$, ${D_{3,8}}$, ${D_{1,4}}$, ${D_{2,5}}$, ${D_{4,20}}$, ${D_{2,16}}$ and ${D_{1,6}}$. ???????????????????????? ????8???12,????????D38???D25?,?????????????D38?????,???1?5????????1{1,4,6}, 2{2,3,5,15,16}, 3{8,10,12,14,18}, 4{19,20}, 5{7,9,11,13,17},???2????{2,5,3,15,16}.
After the instance $(U,{\mathcal{D}},k)$ is input and all dual variables are set to zero, the algorithm enters the while loop. By continuously increasing the dual variables, disks $D_{3,10}$ become tight first. According to the iteration sequence, the disks selected for each iteration are ${D_{3,10}}$, ${D_{4,19}}$, ${D_{5,7}}$, ${D_{3,8}}$, ${D_{1,4}}$, ${D_{2,5}}$, ${D_{4,20}}$, ${D_{2,16}}$ and ${D_{1,6}}$. In this process, each user determines which server serves it. For example, although users $u_8$ and $u_{12}$ are contained in disk $D_{3,8}$ and disk $D_{2,5}$, respectively, the first selected disk, that is, $D_{3,8}$, covers them. Finally, the users served by servers $s_1$ to $s_5$ are $\{u_1, u_4, u_6\}_{s_1}$, $\{u_2, u_3, u_5, u_{15}, u_{16}\}_{s_2}$, $\{u_8, u_{10}, u_{12}, u_{14}, u_{18}\}_{s_3}$, $\{u_{19}, u_{20}\}_{s_4}$, and $\{u_7, u_9, u_{11}, u_{13}, u_{17}\}_{s_5}$.

\subsection{Theoretical Analysis}
\begin{Lemma} \label{lemma1}
	$D_{su}$ must have the capacity to cover all users in $D'_{su}$ when it is selected. Formally, $\left| {D{'_{su}}} \right| \le {k_s}$ when $D{_{su}}$ is selected.
\end{Lemma}

\begin{proof}
	Assume that the $D_{su}$ selected in an iteration satisfies $\left| {D{'_{su}}} \right| > {k_s}$. At this point, $k_s\beta_{su}=p_{su}$, with $\gamma_{h, su}=0, h \in D_{su}$. Then, there must be an unselected disk $D_{sd}, r_{sd} \prec r_{su}$ with $\left| {D{'_{sd}}} \right| = {k_s}$. We have that
	\begin{align}
		{p_{sd}} &\ge {k_s}{\beta _{sd}} + \sum\nolimits_{h:h \in {D_{sd}}} {{\gamma _{h,sd}}} \notag\\
		\label{lemma1:eq2}&= \sum\nolimits_{h:h \in {D'_{sd}}} {{(\beta_{sd} + \gamma _{h,sd})}} \\
		&= \sum\nolimits_{h:h \in {D'_{sd}}} {{\theta_h}} \notag\\
		\label{lemma1:eq4}&= k_s\beta_{su} \\
		&= p_{su} \notag\\
		\label{lemma1:eq6}&> p_{sd}.
	\end{align}
	Equation (\ref{lemma1:eq2}) holds because $\gamma_{h, sd}=0, h\in D_{sd}\backslash D'_{sd}$. Based on the former assumption and Definition \ref{d1}, ${D'_{sd}} \subseteq {D'_{su}}$. Then, $\theta_h=\beta_{su}=(\beta_{sd} + \gamma_{h, sd}), h\in D'_{sd}$. Therefore, Equation (\ref{lemma1:eq4}) holds. In summary, Inequality (\ref{lemma1:eq6}) breaks constraint (\ref{ip:3.2}), and the above assumption is not tenable.
\end{proof}

\begin{Lemma} \label{lemma2}
	We can charge the power of each selected disk $D_{ij}$ to users in $D'_{ij}$ such that each user $h$ obtains a charge of at most $m\cdot\theta_h$.
\end{Lemma}

\begin{proof}
	Define a disk to be a \emph{low degree} disk if $\left| {{D_{ij}}} \right| \le {k_i}$ when it is selected; otherwise, define the disk as a \emph{high degree} disk. We discuss the charging mechanism for both low-degree and high-degree disks. We use $\delta(D_{ij})$ to denote the set of users charged by $i$.

	Consider a low degree disk $D_{ij}$ with $\beta_{ij}=0$. When $D_{ij}$ is selected, we charge all the users contained in this disk; thus, $\delta(D_{ij}) = D_{ij}$. Since the disk constraint is tight, we have ${p_{ij}} = \sum\nolimits_{h:h \in {D_{ij}}} {{\gamma _{h,ij}}} = \sum\nolimits_{h:h \in \delta ({D_{ij}})} {{\theta _h}} $. Thus, we charge the cost of disk $D_{ij}$ to all users contained in this disk by charging $\theta_h$ to each $h\in\delta(D_{ij})$.

	Now, consider a high degree disk $D_{ij}$; then, at some point in time we have $\left| D'_{ij} \right| = k_i$. At this point, we fix the value of $\beta_{ij}$ and subsequently increase the $\gamma_{h, ij}$ variables and make $\delta(D_{ij})=D'_{ij}$. When this disk is declared open $\left| {D{'_{ij}}} \right| \le {k_i}$, we have that ${p_{ij}} = {k_i}{\beta _{ij}} + \sum\nolimits_{h:h \in {D_{ij}}} {{\gamma _{h, ij}}}$. For the users not in $\delta(D_{ij})$, note that $\gamma_{h, ij}=0$. Hence, ${p_{ij}} = {k_i}{\beta _{ij}} + \sum\nolimits_{h:h \in \delta(D_{ij})} {{\gamma _{h, ij}}} $. Since there are exactly $k_i$ users in $\delta(D_{ij})$, we have ${p_{ij}} = \sum\nolimits_{h:h \in \delta(D_{ij})} {({\beta_{ij}} + {\gamma _{h, ij}})} = \sum\nolimits_{h:h \in \delta(D_{ij})} {{\theta_h}} $. Thus, the cost of disk $D_{ij}$ is charged to all users in $\delta(D_{ij})$.

	Finally, Algorithm \ref{alg:PD} selects up to $m$ disks. Therefore, each user in $C$ can be charged up to $m$ times. Formally, $\left| {\{ {D_{ij}}:h \in \delta ({D_{ij}}),{D_{ij}} \in \widehat {\mathcal{D}}\} } \right| \le m,\;h \in U$
\end{proof}

\begin{Theorem}
	\label{theo:ratio}
	The CMPC-PD algorithm returns an $m$-approximation for the capacitated minimum power cover problem in polynomial time.
\end{Theorem}

\begin{proof}
	\emph{(Approximation ratio):}
	For the cover $\widehat {\mathcal{D}}$ constructed by the algorithm, we show that $\sum\nolimits_{{D_{ij}} \in \widehat {\mathcal{D}}} {{p_{ij}}}  \le f \cdot {\text{OPT}}$, where OPT is the value of an optimum solution to the CMPC problem. Let $Z_{LP}^*$ be the optimal value of the linear programming relaxation of (\ref{ip:2}). It is sufficient to show that $\sum\nolimits_{{D_{ij}} \in \widehat {\mathcal{D}}} {{p_{ij}}}  \le f \cdot (\sum\nolimits_{h \in U} {{\theta _h}}  - \sum\nolimits_{i \in S} {{\mu _i}})$ for the final dual solution $\theta$ and $\alpha$ since by weak duality we know that for any dual feasible solution $\theta$ and $\alpha$, $\sum\nolimits_{h \in U} {{\theta _h}}  - \sum\nolimits_{i \in S} {{\mu _i}}  \le Z_{LP}^*$. Thus, since the LP is a relaxation, $Z_{LP}^* \le {\text{OPT}}$.
	
	According to Lemma \ref{lemma2}, we can charge the cost of each chosen disk $D_{ij}$ to the users in $\delta ({D_{ij}})$ at most $m$ times. Thus, we have that
	\begin{eqnarray*}
		\sum\limits_{{D_{ij}} \in \widehat {\mathcal{D}}} {{p_{ij}}} &=&\sum\limits_{{D_{ij}} \in \widehat {\mathcal{D}}} {({k_i}{\beta _{ij}} + \sum\limits_{h:h \in {D_{ij}}} {{\gamma _{h,ij}}}  - {\mu _i})} \\
		&=&\sum\limits_{{D_{ij}} \in \widehat {\mathcal{D}}} {\sum\limits_{h:h \in \delta ({D_{ij}})} {{\theta _h}} }\\
		&=&\sum\limits_{h \in U} {{\theta _h} \cdot \left| {\{ {D_{ij}}:h \in \delta ({D_{ij}}),{D_{ij}} \in \widehat {\mathcal{D}}\} } \right|} \\
		&\le&  m \cdot \sum\limits_{h \in U} {{\theta _h}} \\
		&\le&  m \cdot {\text{OPT}}
	\end{eqnarray*}
	where the second equality is derived from Lemmas \ref{lemma2} and ${\mu _i}{\text{ = }}0, i \in S$.

	\emph{(Polynomial Running Time):} The \textbf{while} loop iterates at most $n$ times to cover all users. Line 4 takes $O(mn^2)$ time to increase the dual variables. Lines 5-7 update several sets in $O(n)$ time. Thus, the CMPC-PD algorithm runs in polynomial time $O(mn^3)$.
\end{proof}

\section{Experimental Results}
We use the PD algorithm proposed above and synthetic data to conduct practical experiments to simulate the power control for servers in edge networks. The experiments ignore the vertical distribution of the two facilities, mapping their positions to a 2-dimensional coordinate system. The relevant parameters are shown in Table \ref{table:param}. The specific experimental settings are as follows:

\begin{enumerate}
	\item The hardware configuration of the experimental environment is as follows: the CPU is an Intel i7-10700 with 8 cores and 16 threads at 2.9 GHz, 16 GB memory, and a hard disk capacity of 1 TB.
	\item The entire system includes two facilities, namely, servers and users, which are distributed randomly in a 2-dimensional space.
	%EDITOR: Please ensure that the intended meaning has been maintained in this edit.
	The capabilities of each server are limited, and all users must be covered by some server. In this experiment, we randomly set the capacity of each server according to its average capacity $\bar k$; thus, the capacity of each server varies around $\bar k$. To ensure that the total system capacity satisfies the requirement of covering all users, if $\sum\nolimits_{i \in S} {k_i} < n$, we artificially increase the gap 
	so that $\sum\nolimits_{i \in S} {k_i}  = n$; if $\sum\nolimits_{i \in S} {{k_i}}  \ge n$, no operation is performed.
	\item The experimental data are generated in an average distribution over a given range, so each experiment is repeated 50 times. And the final results are averaged to reduce the impact of randomness.
	\item The IP utilization rate of a server is the ratio of the number of users covered by it to its capacity.
	\item The variance in the IP utilization is defined by Equation (\ref{eq:variance}):
	\begin{equation}
		{s^2} = \frac{{\sum\nolimits_{i \in S} {{{\left( {\left| {{L_i}} \right| - n/m} \right)}^2}} }}{m}.
		\label{eq:variance}
	\end{equation}
	According to the property of variance, the smaller the value of ${s^2}$ is, the more balanced the number of users covered by each server.
	\item In this section, we compare the PD algorithm with the 
	%Editor: Please ensure that the intended meaning has been maintained in the following edit.
	optimal	and nearest capable server (NCS) approaches. The OPT approach uses IBM's open source tool CPLEX to obtain the optimal solution to the CMPC problem. 
	%Editor: Please ensure that the intended meaning has been maintained in the following edit.
	If we do not obtain the optimal solution within 10 minutes, we stop CPLEX. 
	The NCS approach is a greedy-based method that can be used to solve the CMPC problem. To determine which server covers which user during each iteration, the algorithm selects the closest server-user point pair for which the server still has capacity.
\end{enumerate}

\begin{table}[]
	\caption{Configuration of experimental parameters.}
	\centering
	\begin{tabular}{lll}
		\hline
		\textbf{Param} & \textbf{Description} & \textbf{Value} \\ \hline
		$\alpha$ & Power parameter in Equation (\ref{eq:power}) & [1, 2] \\
		$c$ & Power parameter in Equation (\ref{eq:power}) & 1 \\
		$m$ & \# of servers & [1,10] \\
		$n$ & \# of users & [20,500] \\
		$\bar k$ & Average capacity of all servers & [0,200] \\
		$k_i$ & Capacity of server $i$ & [0,200] \\
		$K$ & Total capacity of all servers & $m\cdot \bar k$ \\
		$l$ & Side length of facility distribution area & 100 \\
		$\lambda$ & Ratio of the side length of the server distribution area to $l$ & [0, 1] \\
		%$\phi$ & ratio of the number of middle servers to $m$ & [0,1] \\
		$p^X_{i}$ ($p^Y_{i}$) & X (Y) coordinate of server $i$ & [0, 100] \\
		$p^X_{j}$ ($p^Y_{j}$)& X (Y) coordinate of user $j$  & [0, 100] \\ \hline
		\label{table:param}
	\end{tabular}
\end{table}

\subsection{Impact of the Number of Users}
\label{sec:ex:n}
\begin{figure}[!t]
	\centering
	\subfigure[Total power]{
		\label{fig:case1:power}
		\includegraphics[width=2.5in]{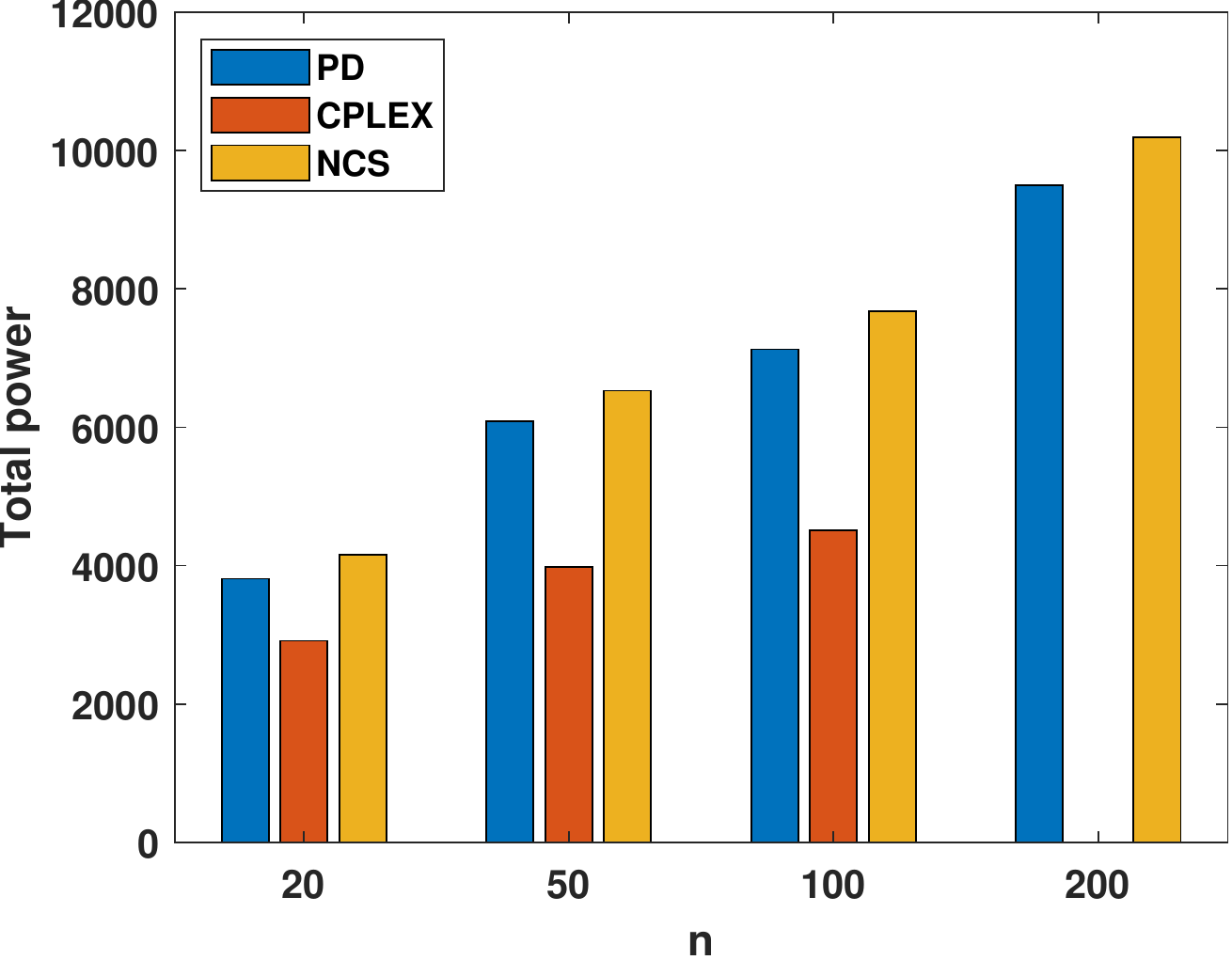}
	}
	\subfigure[Execution time (ms)]{
		\label{fig:case1:time}
		\includegraphics[width=2.5in]{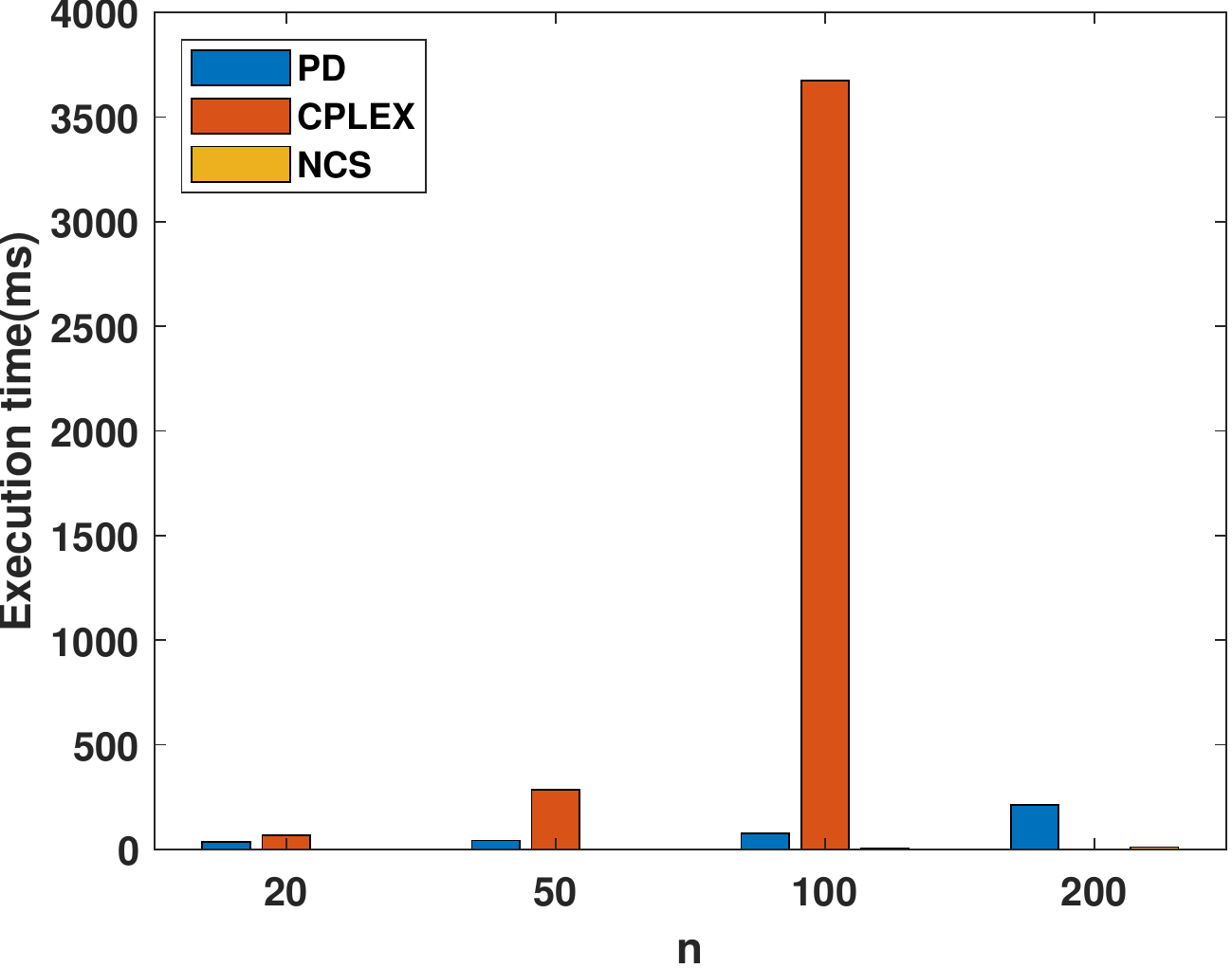}
	}
	\caption{System performance with different numbers of users.}
	\label{fig:case1}
\end{figure}

In this experiment, we analyzed the impact of changes in the number of users on the CMPC problem. The main foci are the total system power, algorithm execution time, and variance in the server's capacity utilization. $n$ is gradually increased from 20 to 200. 
All facilities are distributed in an area with a side length of 100. The average capacity of each server is $\bar k = 50$. Therefore, the total capacity of all servers is $K=500$. For the two constants in Equation (\ref{eq:power}), $c=1$ and $\alpha {\text{ = 2}}$. When the number of users is greater than 200, the optimal solution of a single instance cannot be obtained within 10 minutes. So, the power and execution time of CPLEX seems to be 0 when n=200 in Fig.\ref{fig:case1}.

% ?1??????????,??????????? ????,??????????????????????????????????,???????????????,???????????????????????????????????????????????,???????????????????????????????????????????????,?????DP????NCS?????????????????????????????????????,??2??????????????????????????,OPT??????????
Fig. \ref{fig:case1:power} shows the variation in the total power as the number of users increases for the three approaches. Overall, the total power obtained by the three methods increases as the number of users increases. This is consistent with our intuition. For example, consider several servers installed around a shopping mall. It is certain that the number and capacity of these servers does not change in one day. However, users in the mall change over time. When the number of consumers increases, the probability that the server serves more distant customers increases. The statistics of the total power should then increase. In addition, the DP result is closer to OPT than the NCS result. This finding indicates the superiority of our method. Although our results are not as good as the optimal solution, Fig. \ref{fig:case1:time} shows the considerable time cost required to arrive at the optimal solution.

\subsection{Impact of the Number of Servers and $K$}
\label{sec:ex:m_K}
%  ????,??????????????K??????????? ??????????????????,?????????????????????????? ????????100???,m?K???1???15???100???200?????????1????????

% ??2??????????,???????????(K>=150),????????m????????????????????????????????????,??????????PD???????????????????????2??3?????,??????????????????????????,???????????,????????????????????????,????????,PD??????????,???3???,??3?????????????,????????PD???????????????????3????(????)?
In this section, we study the impact of the number of servers and the total capacity $K$ on the algorithm. This investigation helps us to determine whether to reduce the total power of the system by increasing the number or capacity of the servers when the number of users is stable. In this experiment, we assume that there are 100 users, and $m$ and $K$ increase from 1 to 8 and 100 to 200, respectively. The values of the other variables are the same as those in Section \ref{sec:ex:n}.

\begin{figure}[!t]
	\centering
	\subfigure[Total power]{
		\centering
		\label{fig:case2:power}
		\includegraphics[width=2.5in]{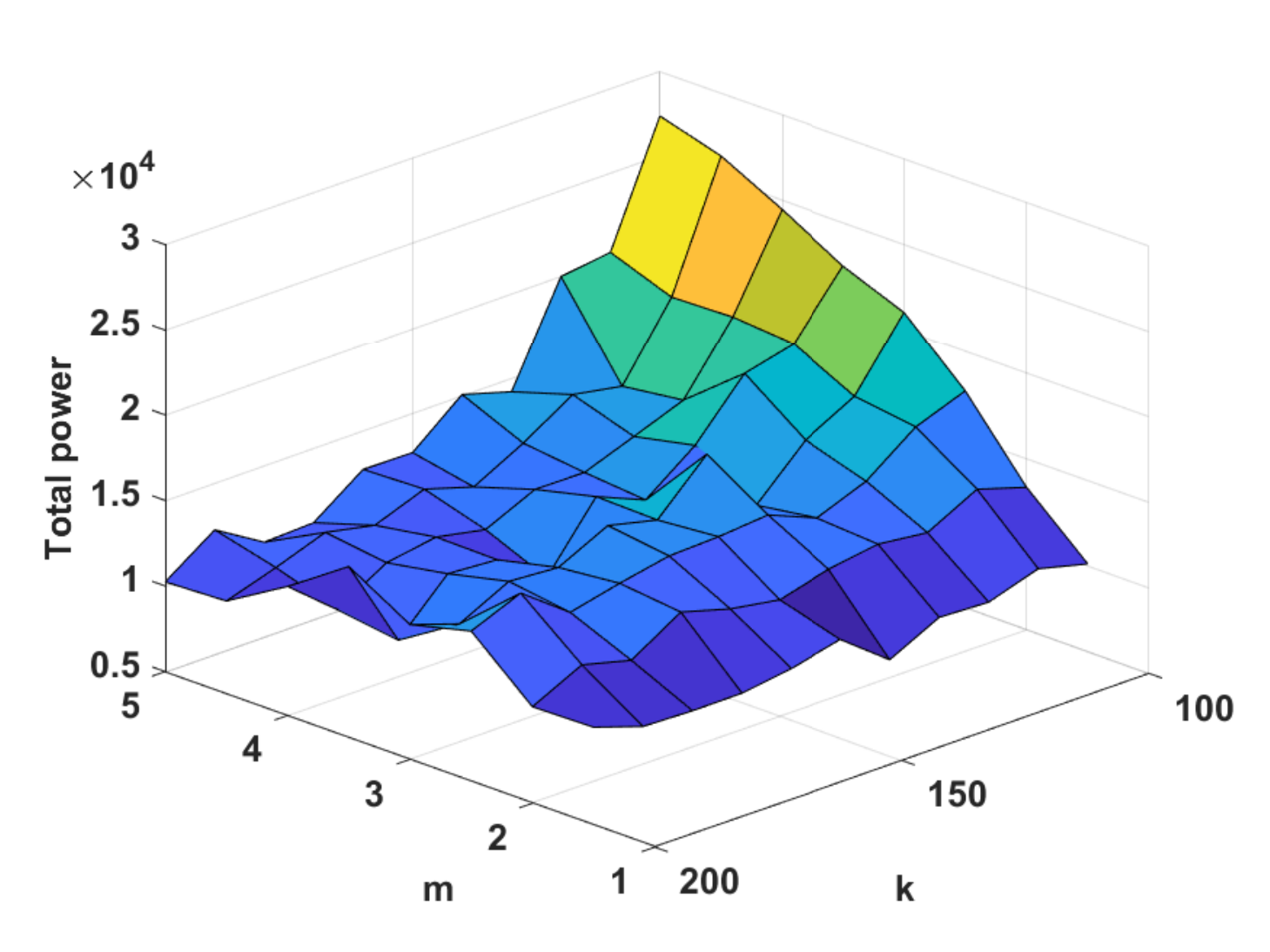}
	}
	\subfigure[Approximation ratio]{
		\centering
		\label{fig:case2:ratio}
		\includegraphics[width=2.5in]{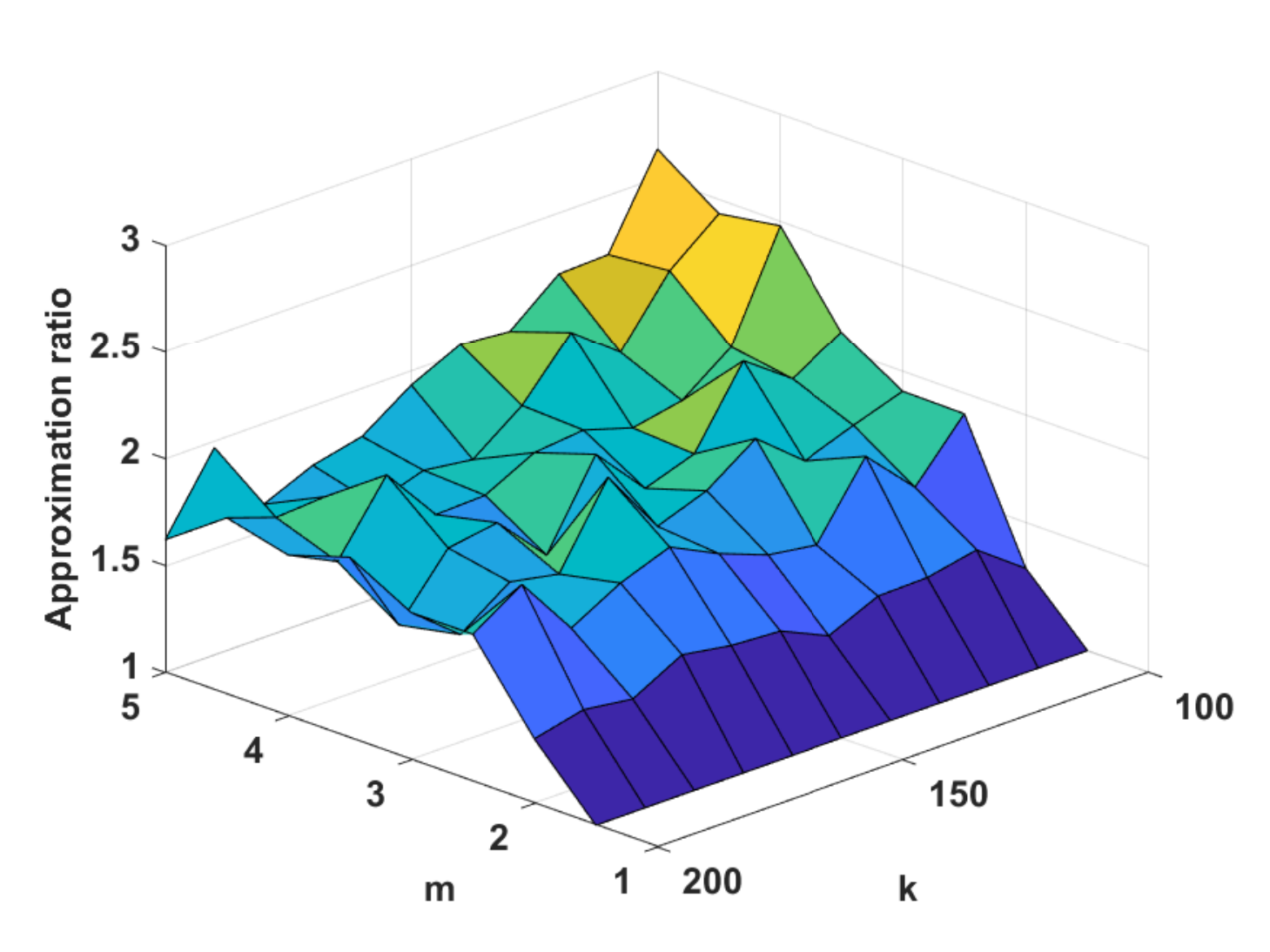}
	}
	\caption{System performance with different numbers of servers and values of $K$.}
\end{figure}

Fig. \ref{fig:case2:power} clearly shows that when the total capacity of the system is sufficient ($K \ge 150$), the total power does not change significantly with increasing $m$. This result indicates that sufficient capacity generates less power. When the system capacity is tight, increasing the number of servers causes the PD algorithm to encounter the issue of local optimization. Combined with the results in Fig. \ref{fig:case2:power} and Fig. \ref{fig:case2:ratio}, this finding leads to a rapid increase in the total power. However, when considering the cost of the server itself, with the assumption of a stable number of users, we can use fewer servers. In this case, the PD algorithm shows good performance regardless of resource constraints, as shown in Fig. \ref{fig:case2:ratio}. In addition, Fig. \ref{fig:case2:ratio} shows that as the number of servers increases, the approximate ratio of the PD algorithm increases. This result confirms the conclusion of Theorem \ref{theo:ratio}.

%  。。。在引言中，我们假设一个服务器可以服务每个用户。如果用户与服务器的分布规律是随机的，在最优解中服务器都会服务相对更近的用户，几乎不可能服务距离最远的用户。本文提出的方法会因局部最优的困境导致某个服务器去服务代价更高的用户，这也是所有近似算法面临的挑战。在图3b中我们可以看到，本文方法只有在少数情况下近似比会超过2，m=[5,4],k=[100,120]这几个少数情况。当近似比小于2时，本文方法得到的圆盘的平均半径不会超过最优解的2^(1/2)倍。
In the introduction, we assume that a server can serve every user. In an optimal solution, a server serves the relatively close user, and rarely the user who is farthest away. The approach proposed in this paper leads to servers serving more costly users due to local optimality, which is a challenge for all approximation algorithms. In Fig. \ref{fig:case2:ratio} we can see that the approximation ratio will only exceed 2 in a few cases where $m \in \{4,5\}$ and $k \in \{100,110,120\}$. When the approximation ratio is less than 2, the average radius of the disk obtained by our approach does not exceed ${2^{\frac{1}{\alpha}}}$ times the optimal solution.

\subsection{Impact of the Number of Servers and $\lambda$}
\begin{figure}[!t]
	\centering
	\subfigure[Total power]{
		\centering
		\label{fig:case3:power}
		\includegraphics[width=6cm]{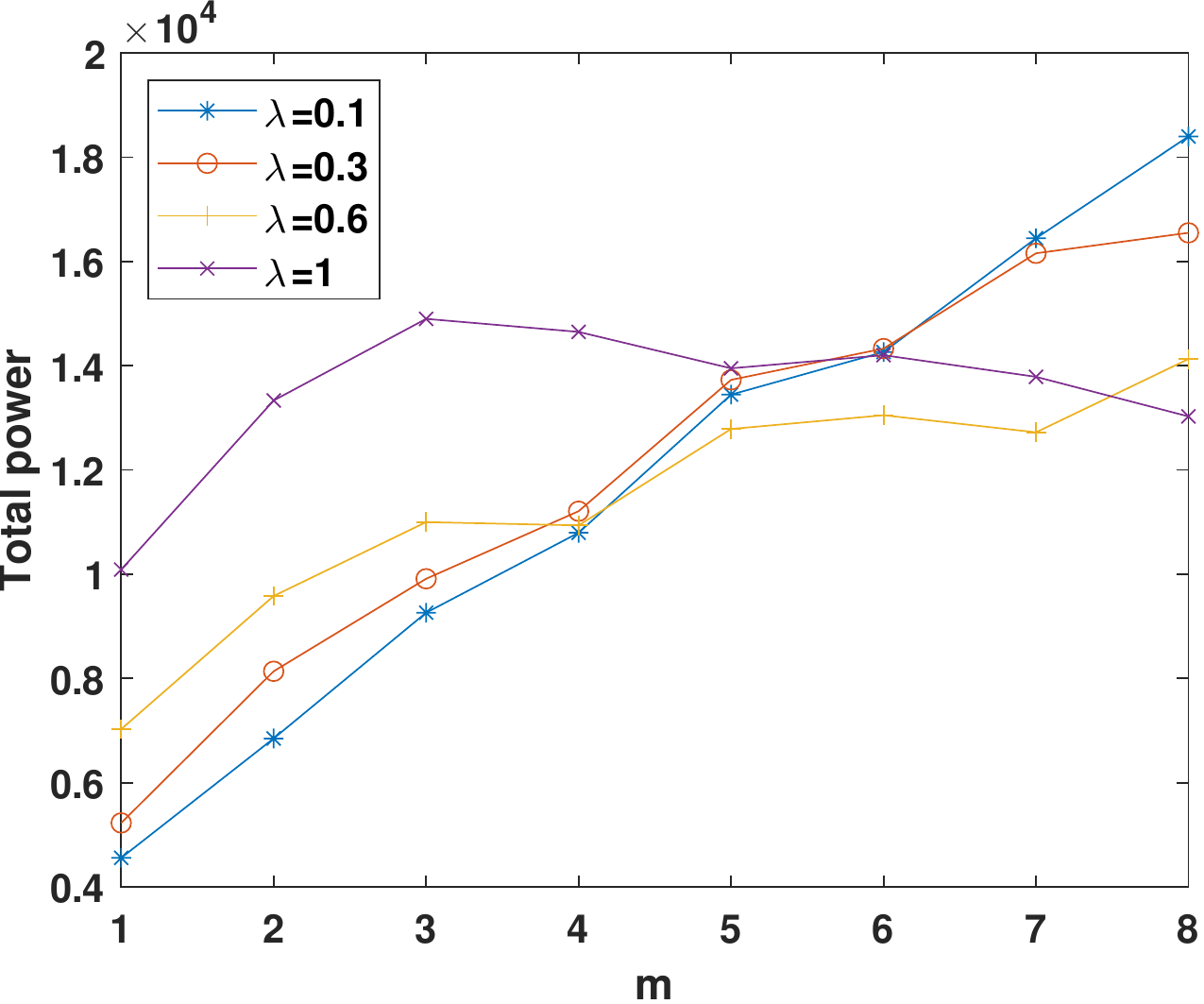}
	}
	\subfigure[Approximation ratio]{
		\centering
		\label{fig:case3:ratio}
		\includegraphics[width=6cm]{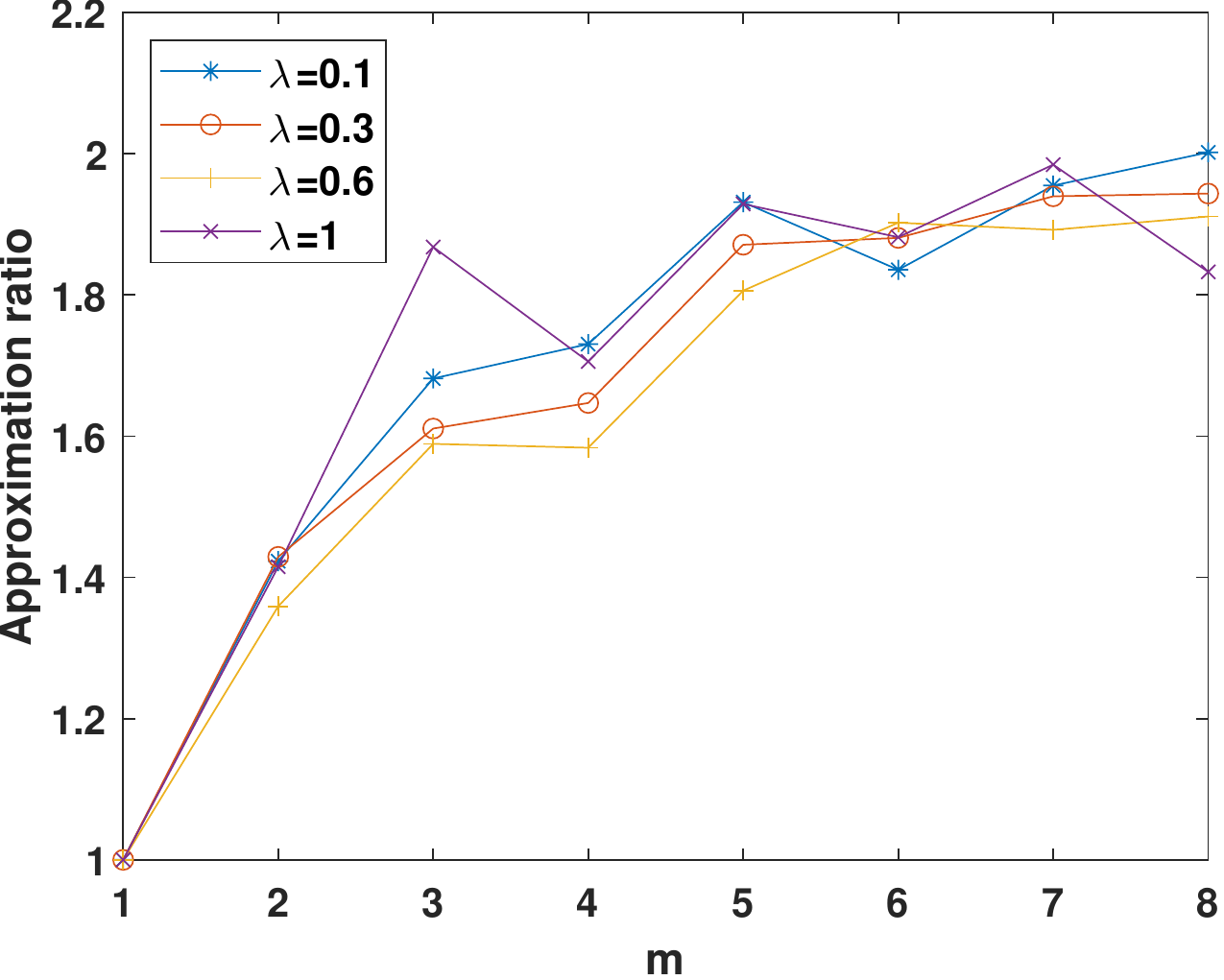}
	}
	\caption{System performance with different numbers of servers and values of $\lambda$.}
\end{figure}
% ??????????????????????????????????????????,??????????????,?????????PD????????????????????l=100?squre?,?lambda??????????,???????lambda*l,????(l/2,l/2)??????????1???8,K=150?????????1????????

% ??4????????,?????????3?,?????????,???????????????????,???????????????????????????8?,??????????????????????????????????????????,????????????????????????????????????????????????,????????????????????????????

Server location problems are also an important research topic. Although this paper does not discuss how to determine the location of the server, we can explore the impact of the server distribution on the PD algorithm by controlling the concentration of the servers. We assume that the two facilities are distributed in a square with a side length of $l = 100$ and use $\lambda$ to control the distribution area of the server. The side length of this area is $\lambda l$, and the center of this region is located at $(l/2, l/2)$. The number of servers $m$ is increased from 1 to 8, with $K = 150$ and $n = 100$. The values of the other variables are the same as those in Section \ref{sec:ex:n}.

Fig. \ref{fig:case3:power} shows that when the number of servers is less than or equal to 3, the more concentrated the servers are, the lower the total power is. However, as the number of servers gradually increases, different concentrations lead to power changes. When the number of servers is increased to 8, the total power ranking result is opposite to the previous result. 
%Editor: Please ensure that the intended meaning has been maintained in the following edit.
Thus, when we need to place servers in an area, to meet the requirement of the system, we need to place servers only near the center of the area; however, if the QoS of users is considered, the servers should be evenly placed in this area rather than only close to the center.

\subsection{Impact of Different Values of $\alpha$}
% ???1?,alpha????????????????????????????????????????????????alpha??????????,???PD???????????????,????alpha???????????,????????????m=6,K=150????

% ?6a????????alpha????????,???PD??CPLEX????????????????????????alpha???PD??????????????????????????b??????,PD???????????alpha????????????alpha??????????????,?variance of IP utilization????alpha????????????????????????????

In Equation (\ref{eq:power}), $\alpha$ is an important parameter that represents the signal attenuation coefficient. Variations in $\alpha$ inevitably affect the total power. Therefore, in this section, we explore the impact of $\alpha$ on the total power and the performance of the PD algorithm. In this experiment, the same dataset was used for different values of $\alpha$. This dataset was developed according to the case when $m = 6$ and $K = 150$, as described in Section \ref{sec:ex:m_K}.

\begin{figure}[!t]
	\begin{adjustwidth}{-\extralength}{0cm}
	\label{fig:case5}
	\centering
	\subfigure[Total power]{
		\label{fig:case5:power}
		\includegraphics[width=2.2in]{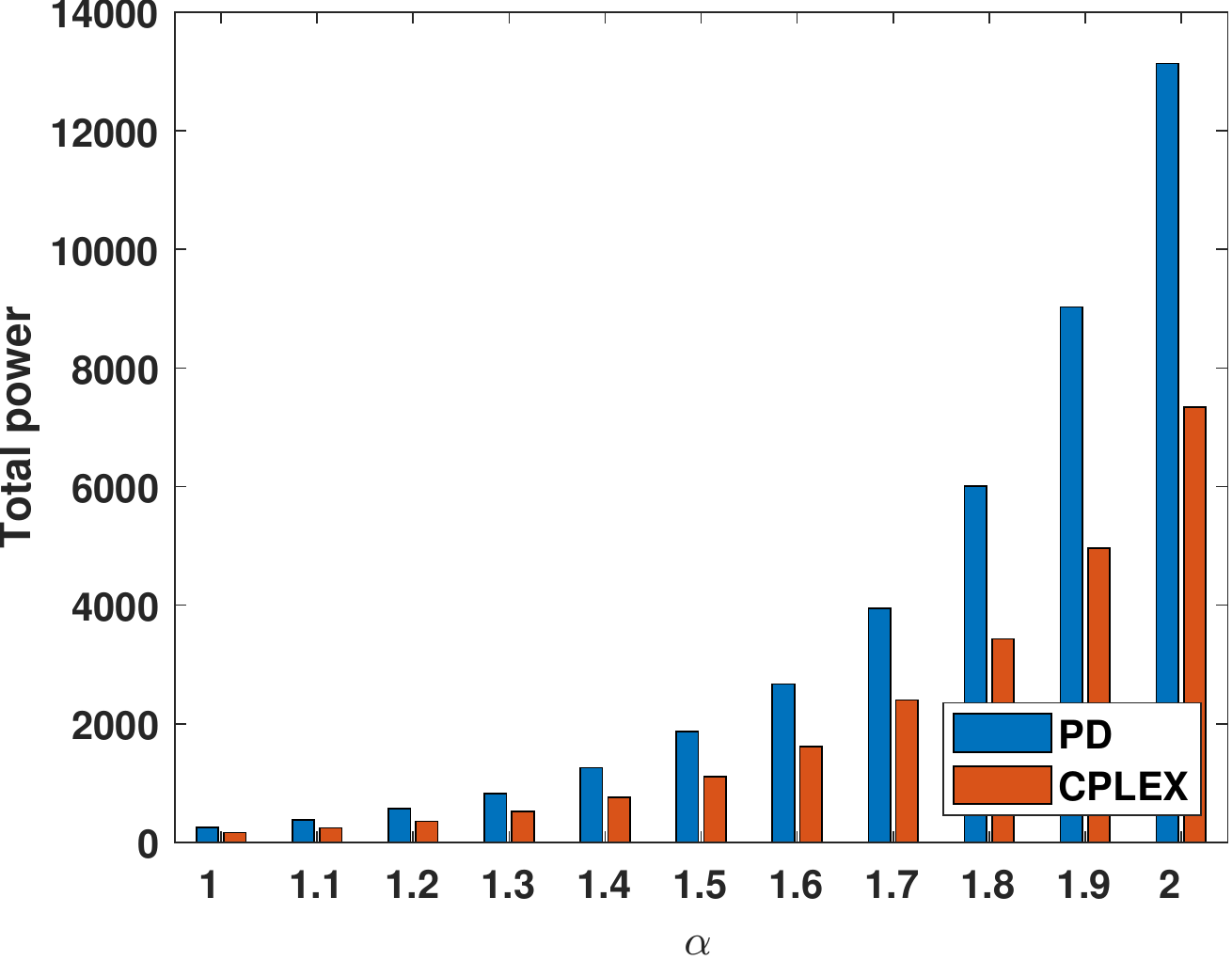}
	}
	\subfigure[Approximation ratio]{
		\label{fig:case5:ratio}
		\includegraphics[width=2.2in]{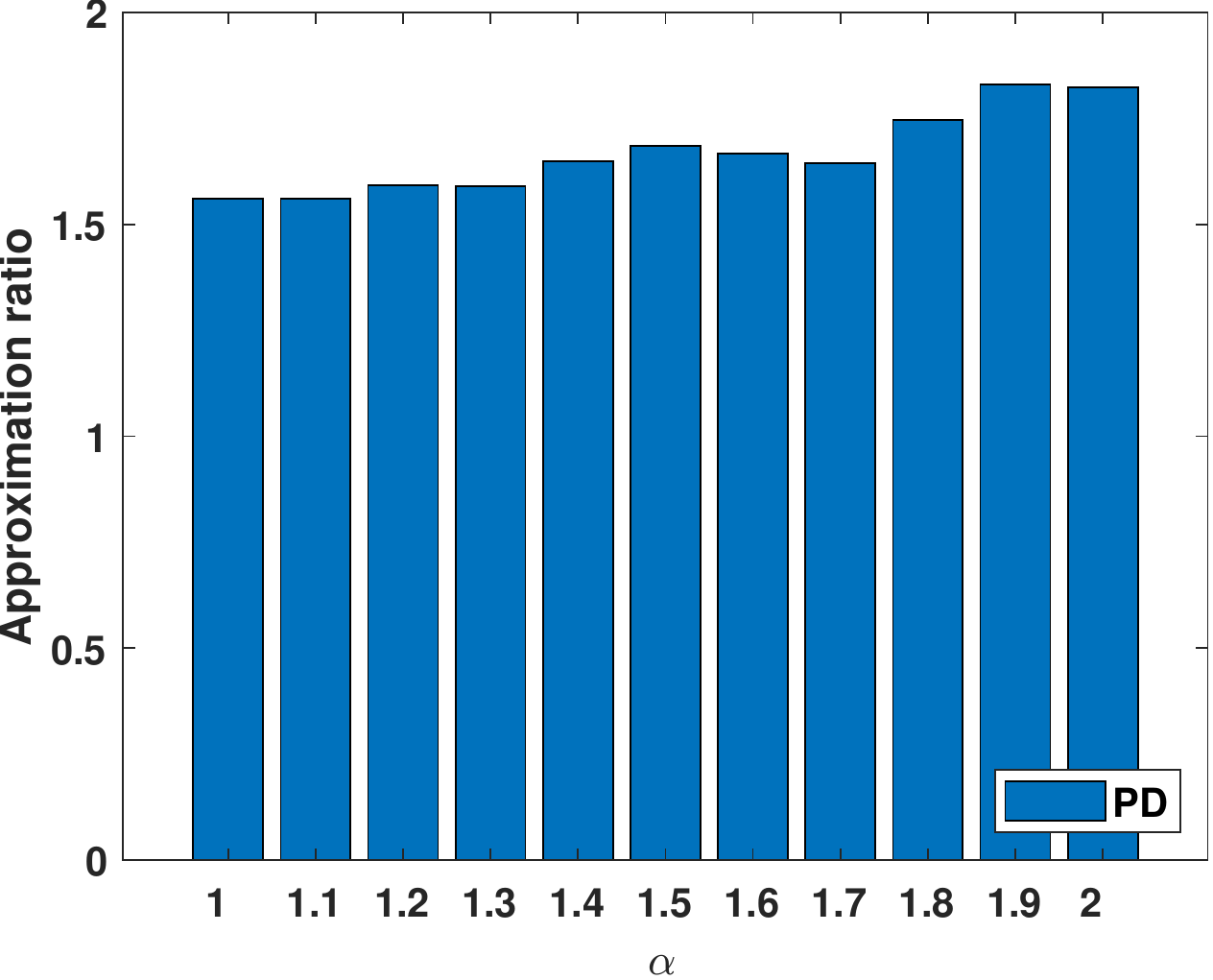}
	}
	\subfigure[Variance in IP utilization]{
		\label{fig:case5:variance}
		\includegraphics[width=2.2in]{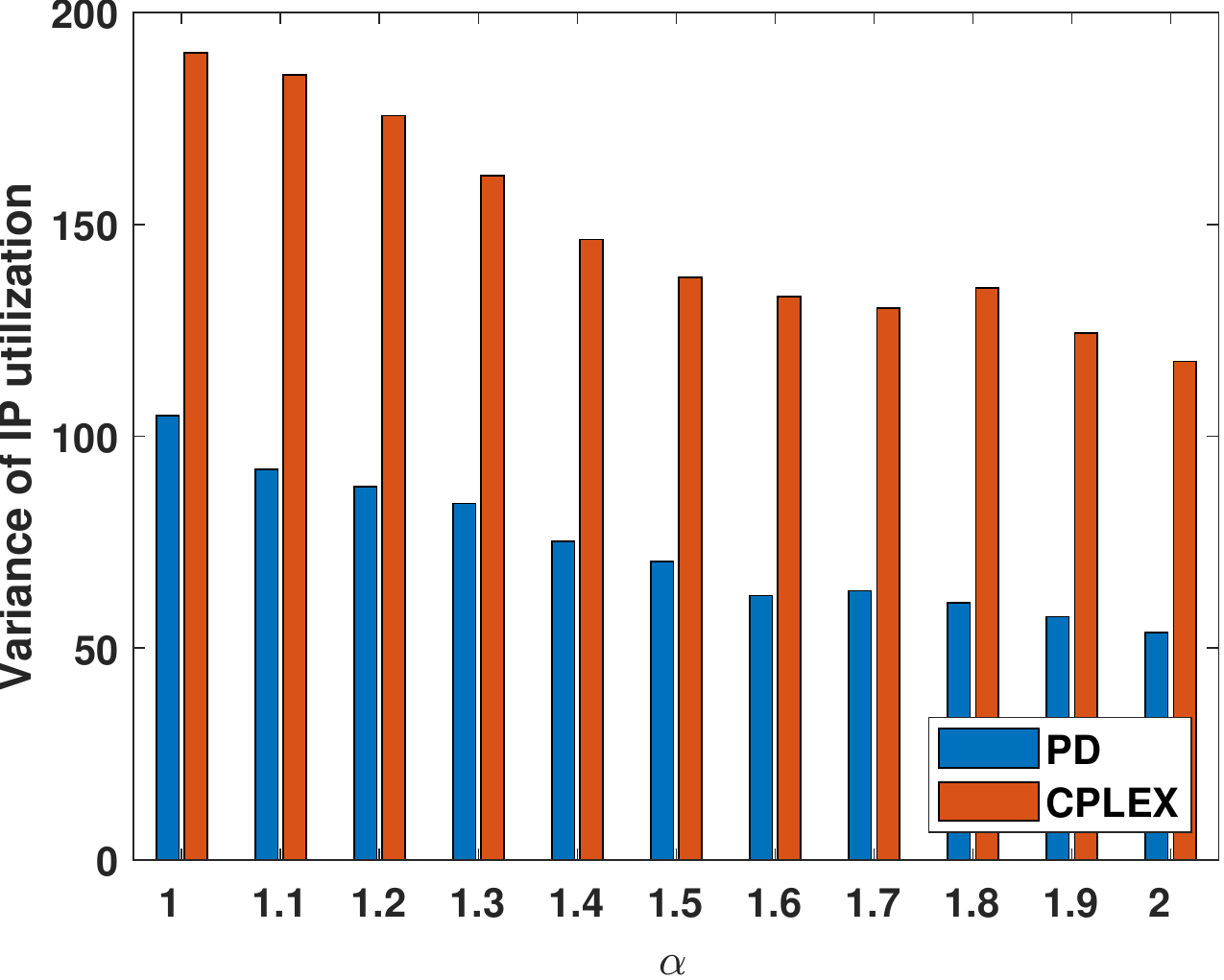}
	}
	\end{adjustwidth}
	\caption{System performance with different values of $\alpha$.}
\end{figure}

Fig. \ref{fig:case5:power} shows that the total power of the PD and CPLEX algorithms increases exponentially with increasing $\alpha$. The reason for this result is clear. A larger value of $\alpha$ increases the cost of the PD algorithm choosing a disk that differs from the optimal solution. 
%Editor: Please ensure that the intended meaning has been maintained in the following edit.
Therefore, as shown in Fig. \ref{fig:case5:ratio}, the approximation ratio of the PD algorithm does not increase with small increases in $\alpha$. 
Although the influence of $\alpha$ on the approximation ratio is not obvious, the variance in the IP utilization decreases with increasing $\alpha$. This result shows that users are increasingly evenly served by the server.

\subsection{Evaluating the gap between assumptions and practical scenarios}

% 在intruduction节中，我们提出了一个变量名为SNR_min，当服务器与用户之间的SNR大于SNR_min，该服务器就能为该用户提供服务。但这个假设与真实情况还有一些差距，我们需要对这个差距进行进一步评估，评估不仅限于4.2小节的分析。
We propose a variable named $SNR_{min}$ in the introduction section. When the SNR between server $s$ and user $u$ is greater than $SNR_{min}$, $s$ can serve $u$. However, there is still some gap between this assumption and the practical scenarios. We need a further evaluation of this gap, not confined to the analysis at the end of subsection 4.2.

%我们假设所有服务器天线的默认覆盖半径r*都相同。这个r*由rou \in {0.5, 0.6,0.7,0.8,0.9,1}确定，rou被定义为服务器半径为r*时能覆盖的用户比例。例如，当rou=0.5时，r*的大小等于服务器信号圆盘刚好能覆盖50%用户时圆盘的半径。本实验中，m=4，K={100~200}，rou={0.5~1}。此外，我们用Ebar来评估the gap between assumptions and actual scenarios。具体定义如下Ebar=。。。
We assume the default coverage radius $r^*$ is the same for all server. The radius $r^*$ is determined by $\rho$, which is defined as the proportion of users that can be covered when the radius of servers is $r^*$. For example, when $\rho=0.5$, $r^*$ is equal to the radius of the server signal disk when it can just cover 50\% of the users. In this experiment, $m=4$, $K$ and $\rho$ increase from 100 to 200 and 0.5 to 1, respectively. In addition, we use $\overline E $ to evaluate the gap between assumptions and practical scenarios, which is defined as follow
\begin{equation}
	\overline E  = \frac{{\sum\nolimits_{i \in S} {\frac{{({r_i} - r^*)}}{{r^*}}} }}{m}, \label{e_gap:eq1}
\end{equation}
where $r_i$ is the radius of server $i$ from the solution of Algorithm 1. As defined in Equation (\ref{e_gap:eq1}), $\overline E$ indicates the average expansion of the radius of the solution obtained by our approach compared to $r^*$.

\begin{figure}[H]
	\begin{adjustwidth}{-\extralength}{0cm}
		\centering
		\includegraphics[width=7cm]{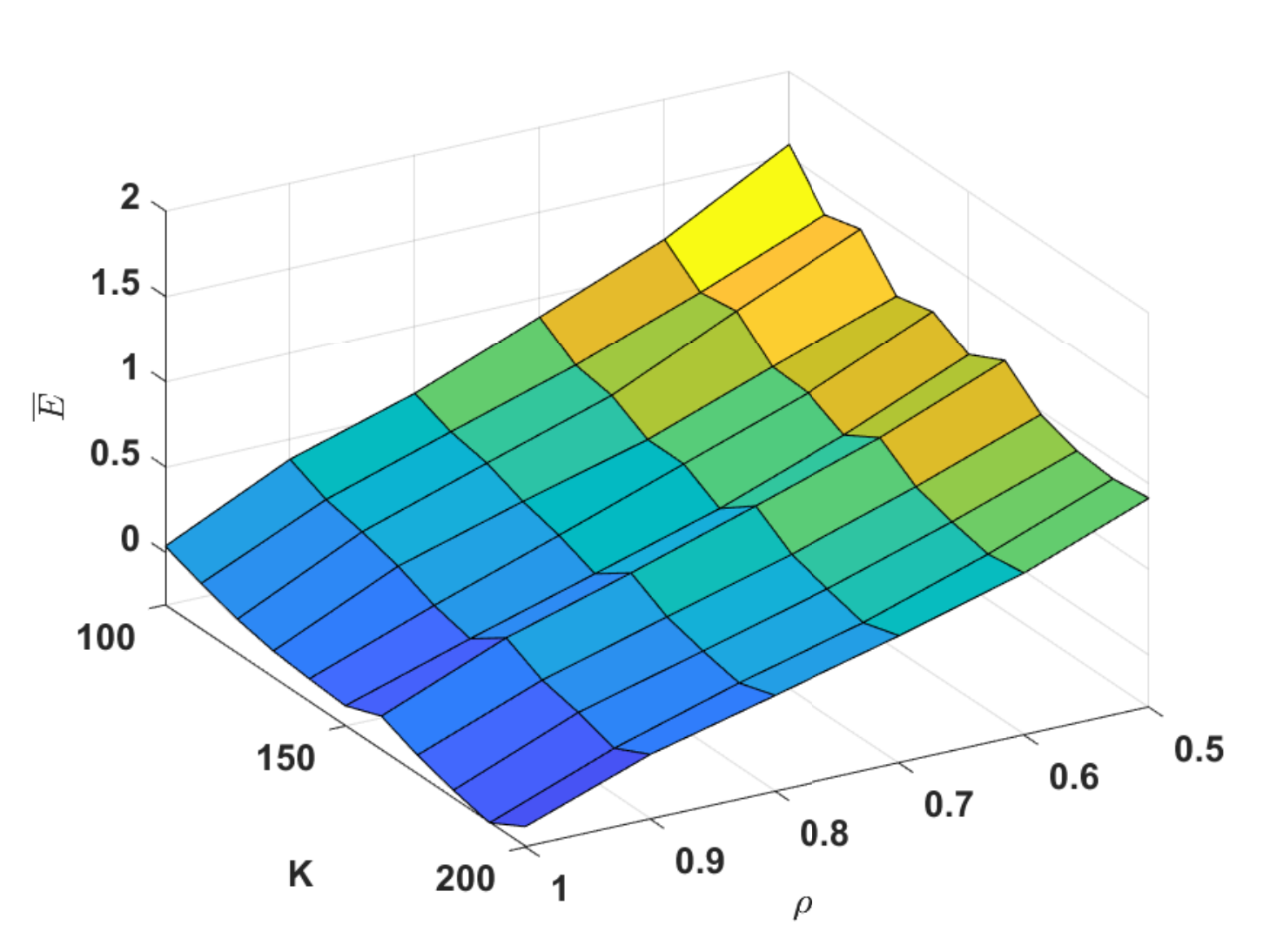}
	\end{adjustwidth}
	\caption{This is a wide figure.\label{fig:case6:E_bar}}
\end{figure}

As we can see in Fig. \ref{fig:case6:E_bar}, if $\rho \geq 0.7$ such that $\overline E \le 1$. It indicates that the server signal disk does not need to be increased by more than a factor of 1 radius when by default, the server can cover more than 70\% of the users. Under the line-of-sight channel model, we think this expansion is acceptable. However, $\overline E$ becomes impractical when $r^*$ is close to 0.5, or the total capacity is very constrained. We can get recommendations on whether additional servers should be established from this result.
%从图6我们可以看出，当

\section{Conclusion}
Signal coverage consumes considerable energy in wireless networks; thus, in this paper, we studied how to assign the appropriate power to servers to reduce energy consumption. We built a signal coverage model based on the CMPC problem in edge networks and developed an $m$-approximation primal-dual-based algorithm. The numerical results show satisfactory performance.

In this paper, we considered that all users need to be covered by servers. 
However, servers need not serve users because of high energy costs, poor communication quality, etc. Therefore, server blocking probability, packet loss, and throughput metrics should focus on future work. These considerations can affect the decision to choose the optimal connecting station for the wireless user. In addition, the power allocation problem for each user after power control for servers would be a further work. Thus, the signal interference between users has to be considered.
% 此外，本文中我们是对每个服务器的天线进行整体的功率控制。
% 在 Practical systems中, 

%%%%%%%%%%%%%%%%%%%%%%%%%%%%%%%%%%%%%%%%%%
\authorcontributions{
	Conceptualization, Qinghui Zhang, Weidong Li and Qian Su; Formal analysis, Qinghui Zhang and Weidong Li; Funding acquisition, Qinghui Zhang and Xuejie Zhang; Methodology, Qinghui Zhang and Weidong Li; Software, Qinghui Zhang; Supervision, Weidong Li and Xuejie Zhang; Validation, Qinghui Zhang, Weidong Li and Xuejie Zhang; Visualization, Qinghui Zhang; Writing – original draft, Qinghui Zhang; Writing – review \& editing, Qinghui Zhang, Weidong Li and Xuejie Zhang.
}

\funding{
	This research was funded by the National Natural Science Foundation of China [Nos. 12071417, 61762091, and 62062065] and the 12th Postgraduate Innovation Project of
	Yunnan University [No. 2020294].
}

\institutionalreview{
	Not applicable.
}

\informedconsent{
	Not applicable.
}

\dataavailability{
	Dataset	link: \url{https://github.com/zqh-ynu/PD-based-data}
} 

\acknowledgments{
	The authors would like to thank all of the cited authors, and the anonymous
	reviewers in this article for their helpful suggestions.
}

\conflictsofinterest{
	The authors declare no conflict of interest.
}

%%%%%%%%%%%%%%%%%%%%%%%%%%%%%%%%%%%%%%%%%%
\begin{adjustwidth}{-\extralength}{0cm}

\reftitle{References}
\bibliography{CMPC_PD_Sensors}
\end{adjustwidth}

\end{document}